\documentclass[conference,a4paper]{IEEEtran}
\IEEEoverridecommandlockouts

\usepackage{cite}
\usepackage{amsmath,amssymb,amsfonts}
\usepackage{algorithm}
\usepackage{algorithmic}
\usepackage{graphicx}
\usepackage{textcomp}
\usepackage{xcolor}
\usepackage{array}
\usepackage{booktabs}

\usepackage[caption=false,font=footnotesize]{subfig}

\def\BibTeX{{\rm B\kern-.05em{\sc i\kern-.025em b}\kern-.08em
    T\kern-.1667em\lower.7ex\hbox{E}\kern-.125emX}}
\makeatletter
\def\@IEEEpubidpullup{3\baselineskip}
\makeatother
\begin{document}

\title{An Ensemble Machine Learning Approach for Detecting Intrusion using Advance Feature Selection Technique
}

\author{
\IEEEauthorblockN{
Md Raihan Al Masud\IEEEauthorrefmark{1},
Hossen Asiful Mustafa\IEEEauthorrefmark{2}}
\IEEEauthorblockA{
\IEEEauthorrefmark{1}ETS, Canada,
\IEEEauthorrefmark{2}Bangladesh University of Engineering and Technology, Bangladesh\\
\IEEEauthorrefmark{1}raihan.emasud@gmail.com,
\IEEEauthorrefmark{2}hossen\_mustafa@iict.buet.ac.bd}
}




\IEEEpubid{\begin{minipage}{\textwidth}
\footnotesize
\textcopyright~2026 IEEE. Personal use of this material is permitted. Permission from IEEE must be obtained for all other uses, in any current or future media, including reprinting/republishing this material for advertising or promotional purposes, creating new collective works, for resale or redistribution to servers or lists, or reuse of any copyrighted component of this work in other works.
\end{minipage}}

\maketitle

\begin{abstract}

The growing number of cyberattacks has increased the demand for effective Network Intrusion Detection Systems (NIDSs). Existing detection systems use traditional machine learning algorithms with low detection rates and limited capability to detect new and unknown network attacks. In this paper, we use a bagging-based ensemble machine learning algorithm to achieve a higher detection rate in some cases and reduce the number of features using a feature selection method. Compared with traditional and deep learning classifiers, ensemble machine learning classifiers provide effective detection of network attacks. The main focus of this work is to improve the detection rate of network attacks using a reduced number of features. We use the NSL-KDD training and test datasets to train and evaluate the classifier, respectively. This work consists of two parts. In the first part, after preprocessing, we investigate how the detection rate changes with an increasing number of selected features. In the second part, we evaluate the multiclass performance of the proposed method and compare the results with conventional machine learning and deep learning classifiers in the literature. Existing methods show lower detection rates for attack classes with fewer samples in an imbalanced dataset. To further evaluate the multiclass performance, the proposed classifier is evaluated using precision-recall curves.

\end{abstract}


\section{Introduction}
As network traffic grows rapidly, network attacks become more numerous and diverse. In 1980, J. P. Anderson introduced an early intrusion detection concept based on system audit data to identify unauthorized access and abnormal user activities \cite{ref1}. This work contributed to the development of intrusion detection systems (IDSs). An IDS inspects network traffic for potentially malicious activity. Signature-based IDSs detect attacks using known attack patterns and may fail to detect unknown attacks. In contrast, anomaly-based IDSs analyze traffic behavior and can detect previously unknown attacks. However, the rapid growth of Internet of Things (IoT) devices, cloud-based services, large volumes of network data, and diverse network protocols makes accurate intrusion detection challenging.

Network attack detection has been studied using a range of machine learning (ML) methods \cite{ref2}. Large, imbalanced datasets can reduce the detection performance of these methods, especially for attack classes with few samples \cite{ref3}. To address this issue, researchers have applied various approaches, including traditional ML, deep learning, and ensemble learning. Ensemble methods combine multiple classifiers and can improve intrusion detection performance compared with individual ML models.

\IEEEpubidadjcol
In this paper, we focus on an anomaly-based network intrusion detection system. In our work, we use the ensemble machine-learning technique based on network traffic patterns. We investigate whether a bagging ensemble model can improve detection of attacks represented by few records in a large, imbalanced dataset. When samples are few, we aim to improve detection rates for each network attack type. Few studies improve detection separately for individual attack categories. Most studies focus on improving DoS detection; however, the detection rates for other attacks, e.g., Probe, R2L, and U2R, were not improved in the same way. Therefore, our main focus is on improving the detection rate for different types of attacks one by one.  We validate our model with the help of scikit-learn by analyzing the full training and test datasets \cite{ref4}. Because NSL-KDD serves as a common benchmark in related studies, we can compare our results directly. The contributions of this paper are summarized below.

\begin{enumerate}
    \item We introduce a feature selection technique based on p-values for detecting intrusions by analyzing the significance of every feature.
    \item We apply bagging and boosting ensemble ML techniques and propose a bagging ML model that is based on a base ML classifier.
    \item We propose a reduced number of features to evaluate our model and improve the detection rate in some cases.
   \item Our model is efficient in detecting Probe, R2L, and U2R attacks similarly compared to DoS attacks using a feature selection technique \cite{ref21} \cite{ref22} \cite{ref23} in ensemble ML.
    \item We compare performance with that of RF, SVM, BDB, LSTM, CNN, DeciTree, KNN, and DNN models in the literature that also report detection rates. 
\end{enumerate}

\section{Related work}\label{sec:related}

Researchers have introduced various machine learning techniques to improve intrusion detection strategies. The support vector machine (SVM) was proposed by Kuang et al. in 2014, and they used the kernel trick in combination with a genetic algorithm (GA) \cite{ref6}. To detect intrusions in a wireless sensor network, the K-Nearest Neighbour (KNN) approach was proposed in 2014 \cite{ref7}. Using an Artificial Neural Network (ANN) on NSL-KDD, Ingre and Yadav \cite{ref9} obtained a small improvement in detection rates for the four attack types. Researchers have also applied the Deep Belief Network (DBN) to NSL-KDD \cite{ref11}. Kim et al. \cite{ref12} used Long Short Term Memory (LSTM) in an LSTM-RNN intrusion detection system and obtained improved accuracy. However, their experiments used NSL-KDD only in part. Dimensionality reduction was proposed by Abolhasanzadeh, who used several techniques on the dataset \cite{ref14}. Good classification ability was achieved by Fiore et al. using a Discriminative Restricted Boltzmann Machine (DBM). However, the performance of all the methods is limited.


The CNN-based intrusion detector developed by Ding et al. \cite{ref15} was evaluated against RF, SVM, DBN, and LSTM. Mirza et al. \cite{ref16} combined logistic regression, neural networks, and decision trees to perform binary classification on the KDD Cup 99 dataset. In 2019, Hu et al. \cite{ref17} proposed Dynamic Deep Forest, which combines feature selection and tree-based ensemble classification. To examine individual attack classes in imbalanced datasets, Sharma et al. \cite{ref18} used an ensemble approach with ExtraTree-based feature selection. On NSL-KDD, Gao et al. \cite{ref19} evaluated RF, voting, and MultiTree classifiers alongside KNN, LR, SVM, and DNN. Our earlier study \cite{ref24} applied a majority-voting ensemble technique to detect network attacks.

Gaikwad et al. \cite{ref25}  propose an intrusion detection system (IDS) that uses the Bagging ensemble method with Reduced Error Pruning Tree (REPTree) as the base classifier.  The authors report higher detection accuracy and lower false-positive rates on NSL-KDD compared with individual classifiers. The study emphasizes that ensemble techniques like Bagging offer more robust and reliable intrusion detection for network security. Barach et al. \cite{ref26} combine Convolutional Neural Networks (CNN) and an attention mechanism in an intrusion detection system designed to identify malicious network activity. Using the NSL-KDD dataset, their model achieves high accuracy and recall, demonstrating that the attention-enhanced CNN effectively highlights important input features and improves detection performance in cybersecurity applications. They did not show the recall of different types of attacks (DoS, Probe, R2L, and U2R) separately, which is important for understanding detection performance. Similarly, Mohammadpour et al. \cite{ref27}, Al-Turaiki et al. \cite{ref28}, Sapre et al. \cite{ref29}, Magdy et al. \cite{ref30} and Zakariah et
al. \cite{ref31} reported the overall recall and accuracy.

\begin{table}[t]
\centering
\caption{Detection Rate of Different ML Classifiers}
\label{tab:related}

\begin{tabular}{lcccc}
\hline
Detection classifier & DoS (\%) & Probe (\%) & R2L (\%) & U2R (\%) \\
\hline
CNN \cite{ref15}      & 85 & 73 & 17 & 8 \\
DeciTree \cite{ref17} & 84 & 66 & 21 & 7 \\
R.Forest \cite{ref17} & 81 & 68 & 2  & 1 \\
KNN \cite{ref17}      & 79 & 62 & 5  & 4 \\
LR \cite{ref17}       & 79 & 64 & 1  & 3 \\
SVM \cite{ref17}      & 77 & 48 & 9  & 2 \\
DNN \cite{ref17}      & 84 & 77 & 30 & 26 \\
\hline
\end{tabular}

\end{table}

In our research work, we propose an ensemble machine learning technique called Bagging, and we use the full NSL-KDD dataset to observe and analyze traffic data and finally compare our findings with CNN \cite{ref15}, DeciTree \cite{ref19}, R. Forest \cite{ref19}, KNN \cite{ref19}, LR \cite{ref19}, SVM \cite{ref19} and DNN \cite{ref19} in the literature that show the detection rates of different types of attacks separately. Table~\ref{tab:related} shows the detection rates of related works.

\section{Ensemble Methods}\label{sec:background}
In this section, we discuss bagging and boosting ensemble learning methods.

\subsection{Bagging}\label{subsec4.1}

Bagging, also known as Bootstrap Aggregation, is an ensemble learning technique that trains multiple models using different bootstrap samples of the training dataset. These samples are generated randomly with replacement, and a separate model is trained on each sample. Final predictions are obtained from all model outputs by majority voting in classification and averaging in regression. By combining predictions from multiple models trained on different data subsets, Bagging reduces variance and helps mitigate overfitting, often providing more reliable predictions than a single model.

\subsection{Boosting}\label{subsec4.2}
The learning process of Boosting is sequential. In this model, the algorithms work by training on the entire dataset and the subsequent algorithms are selected based on the residual error of the initial situations. In this situation, the Boosting algorithm assigns a higher weight to the model that was poorly estimated in the previous stage. This process relies on multiple weak learners that are effective for some portions of the dataset but are not effective for the entire dataset. Therefore, every model improves the performance in the ensemble learning process.

\section{Proposed Methodology}\label{sec_method}
This section shows the mechanism we propose. Our mechanism is divided into four main parts:

\begin{enumerate}
    \item  Data Collection and Preprocessing
    \item Feature Selection
    \item Building the Model
    \item Model Evaluation
\end{enumerate}

\begin{figure}[!htbp]
\centering
\includegraphics[
    width=0.98\columnwidth,
    keepaspectratio
]{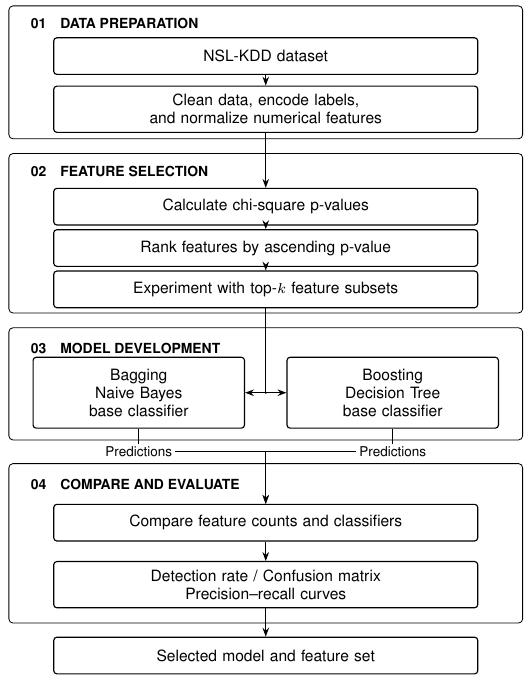}
\caption{Workflow diagram of the proposed technique.}
\label{fig:flow}
\end{figure}

Each stage is divided into several parts. The step-by-step process is discussed below. Once the model is built, we evaluate it using the test dataset.  Fig.~\ref{fig:flow} shows the proposed workflow, and Algorithm 1 shows its steps.



Table~\ref{tab:notation} summarizes the notation used in the proposed method.

\begin{table}[!t]
\centering
\caption{Notation Used in the Proposed Method}
\label{tab:notation}
\footnotesize
\setlength{\tabcolsep}{3pt}
\renewcommand{\arraystretch}{1.02}
\begin{tabular}{@{}p{0.35\columnwidth}p{0.61\columnwidth}@{}}
\toprule
\textbf{Symbol} & \textbf{Meaning} \\
\midrule
$\boldsymbol{\mathcal{D}}$ & Network traffic dataset. \\
$\boldsymbol{\mathcal{D}}_{\mathrm{train}},\boldsymbol{\mathcal{D}}_{\mathrm{test}}$ & Training and test dataset. \\
$m,n,C$ & Numbers of samples, candidate features, and classes, respectively. \\
$\boldsymbol{X}_i,\boldsymbol{y}$ & Feature column $i$ and target-class vector. \\
$\chi_i^2,p_i$ & Chi-square statistic and p-value of feature $i$. \\
$\mathcal{P}$ & Collection of feature-index/p-value pairs. \\
$\mathcal{P}_{\mathrm{sorted}},\mathcal{P}_{\text{top-}k}$ & Ranked pairs and the top-$k$ pairs. \\
$\mathcal{X}_{\mathrm{ranked}}$ & Ranked feature list. \\
$\mathcal{X}_k,\mathcal{X}_{k^{*}}$ & Top-$k$ and selected feature subsets. \\
$x_j$  & Feature at ranked position $j$. \\
$x,x^{\prime}$ & Original and standardized feature values. \\
$x_{mean},\sigma$ & Feature mean and standard deviation. \\
$\mathrm{DR}(k)$ &  detection rate using $k$ features. \\
\bottomrule
\end{tabular}
\end{table}

\subsection{Data Collection and Preprocessing}\label{subsec5.1}

We obtained NSL-KDD from the Canadian Institute for Cybersecurity (CIC) for this study. NSL-KDD addresses several limitations of the original KDD CUP-99 dataset. In this work, KDDTrain+ and KDDTest+ are used, containing 125,973 and 22,544 network traffic samples, respectively.

The dataset contains 43 features, including the ``class'' feature, which represents the traffic class or attack type. We use the class feature as the prediction target and the other 42 features as training inputs. Before training, several preprocessing steps are applied, as described below.

\subsubsection{Data Cleaning}\label{subsec5.2}
After data extraction, the next step is data cleaning. We perform data cleaning using a custom program developed in the Python programming language. By examining the Canadian NSL-KDD network dataset, we remove one of the 42 features because it contains a constant value. We also remove two features which contain non-categorical data. After cleaning, the dataset retains 39 useful features. The 39 useful features are presented in Table~\ref{tab:attr} with their data types.

\begin{table}[!t]
\centering
\caption{Attribute Values and Types}
\label{tab:attr}
\footnotesize
\renewcommand{\arraystretch}{1.15}
\setlength{\tabcolsep}{3pt}

\begin{tabular}{p{0.78\columnwidth} c}
\hline
\textbf{Features} & \textbf{Type} \\
\hline

Protocol\_type (2), Flag (3)
& Nominal \\

\hline

Land (6), logged\_in (11), root\_shell (13),
su\_attempted (14), is\_host\_login (19),
is\_guest\_login (20)
& Binary \\

\hline

Duration (1), src\_bytes (4), dst\_bytes (5),
wrong\_fragment (7), urgent (8), hot (9),
num\_failed\_logins (10), num\_compromised (12),
num\_root (15), num\_file\_creations (16),
num\_shells (17), num\_access\_files (18),
count (21), srv\_count (22), serror\_rate (23),
srv\_serror\_rate (24), rerror\_rate (25),
srv\_rerror\_rate (26), same\_srv\_rate (27),
diff\_srv\_rate (28), srv\_diff\_host\_rate (29),
dst\_host\_count (30), dst\_host\_srv\_count (31),
dst\_host\_same\_srv\_rate (32),
dst\_host\_diff\_srv\_rate (33),
dst\_host\_same\_src\_port\_rate (34),
dst\_host\_srv\_diff\_host\_rate (35),
dst\_host\_serror\_rate (36),
dst\_host\_srv\_serror\_rate (37),
dst\_host\_rerror\_rate (38),
dst\_host\_srv\_rerror\_rate (39)
& Numeric \\

\hline
\end{tabular}
\end{table}

\subsubsection{Data Labeling}\label{subsec5.3}

Three NSL-KDD features are non-numeric: ``protocol\_type, ``flag, and ``class. These categorical values are converted into numeric values using data labeling. TCP, UDP, and ICMP in ``protocol\_type'' are represented by 0, 1, and 2, respectively. Similarly, the 11 categories of the ``flag'' feature are converted into numeric values. The ``class'' feature is used as the target and contains different types of network attacks. These attacks are grouped into five categories: DoS, Probe, R2L, U2R, and Normal, as shown in Table~\ref{tab:mapping}. The four attack categories are briefly described below.

\begin{table}[!t]
\centering
\caption{Mapping of Different Attack Classes With Attack Types}
\label{tab:mapping}
\footnotesize
\renewcommand{\arraystretch}{1.1}
\setlength{\tabcolsep}{6pt}

\begin{tabular}{p{0.30\columnwidth} p{0.40\columnwidth} p{0.12\columnwidth}}
\hline
\textbf{Training Set} & \textbf{Testing Set} & \textbf{Category} \\
\hline

teardrop, smurf, pod, neptune, land, back
&
udpstorm, worm, teardrop, smurf, pod,
neptune, mailbomb, land, back, apache2
&
DoS \\

\hline

satan, portsweep, nmap, ipsweep
&
satan, saint, portsweep, nmap, mscan,
ipsweep
&
Probe \\

\hline

warezmaster, phf, multihop, imap,
guesspasswd, ftpwrite, warezclient, spy
&
xsnoop, warezmaster, xlock, snmpguess,
snmpgetattack, sendmail, phf, named,
multihop, imap, httptunnel, guesspasswd,
ftpwrite
&
R2L \\

\hline

rootkit, loadmodule, ps, buffer\_overflow
&
xterm, sqlattack, loadmodule, perl, ps,
buffer\_overflow
&
U2R \\

\hline

normal
&
normal
&
Normal \\

\hline
\end{tabular}
\end{table}


\textbf{Denial of Service (DoS)} – A DoS attack disrupts the availability of a network or system by consuming its computing and memory resources, preventing legitimate users from accessing services. In the NSL-KDD dataset, common DoS attacks include neptune, smurf, teardrop, pod, and back.
    
\textbf{Probe} – A Probe attack collects information about a target network, such as active hosts, open ports, and vulnerable services, which may be used for future attacks. In the NSL-KDD dataset, common Probe attacks include satan, nmap, ipsweep, and portsweep.

\textbf{Remote to Local (R2L)} – In an R2L attack, an attacker outside the target system exploits vulnerabilities or weak authentication to seek unauthorized access. In the NSL-KDD dataset, common R2L attacks include $guess\_passwd$, $ftp\_write$, phf, imap, warezmaster, and warezclient.
    
\textbf{User to Root (U2R)} – In a U2R attack, an attacker who already has limited user access exploits system vulnerabilities to seek root or administrative privileges. In the NSL-KDD dataset, common U2R attacks include buffer\_overflow, loadmodule, perl, and rootkit.
    

Table~\ref{tab:dataset} shows sample counts for individual traffic classes. The dataset is highly imbalanced. For example, KDDTrain+ contains 67,343 Normal samples but only 52 U2R samples. A similar imbalance is observed in KDDTest+. Such class imbalance makes classification challenging, particularly for detecting minority attack classes.

\begin{table}[!t]
\centering
\caption{Samples of Normal and Attack Data in the Dataset}
\label{tab:dataset}
\footnotesize
\renewcommand{\arraystretch}{1.15}

\begin{tabular*}{\columnwidth}{@{\extracolsep{\fill}}lcc@{}}
\hline
\textbf{Attack Type} & \textbf{KDDTrain+} & \textbf{KDDTest+} \\
\hline
DoS    & 45,927  & 7,460  \\
Probe  & 11,656  & 2,421  \\
U2R    & 52      & 67     \\
R2L    & 995     & 2,885  \\
Normal & 67,343  & 9,711  \\
\hline
\textbf{Total} & \textbf{125,973} & \textbf{22,544} \\
\hline
\end{tabular*}
\end{table}

\subsubsection{Data Normalization}\label{subsec5.4}

Before feature selection, numerical features are normalized using \texttt{StandardScaler()} as in \eqref{eq:normalization}, where $x$ is the feature value, $x_{mean}$ its mean, and $\sigma$ its standard deviation.

\begin{equation}
\label{eq:normalization}
\displaystyle x^{'} \ =\ ( x\ -\ x_{mean}) \ /\ \sigma
\end{equation}




\subsection{Feature Selection}\label{subsec5.4.1}
There are several feature selection techniques available. Among these, the filter method is the fastest. There are some variations among the filter methods. A chi-square test is used in our filter-based method to evaluate how each feature is related to the target variable. To identify the importance of a feature, we evaluate the p-value of a specific feature.
The p-value can be used to select the important features that are statistically significant, and based on these statistical data, we will be able to classify different types of attacks and normal network traffic. The smaller the p-value, the more important the feature is. We use scikit-learn to compute a p-value for every feature in the dataset. The step-by-step procedure is described below: 

\subsubsection{Dataset representation}\ The dataset is represented by \eqref{eq:dataset}:

\begin{equation}
\label{eq:dataset}
 \boldsymbol{\mathcal{D}} = [\boldsymbol{X}_1, \boldsymbol{X}_2, \dots, \boldsymbol{X}_n] \in \mathbb{R}_{\ge 0}^{m \times n}
\end{equation}
where \( \boldsymbol{X}_i \in \mathbb{R}_{\ge 0}^m \) represents the $i$-th feature column with $m$ samples and \( \boldsymbol{y} \in \{1, 2, \dots, C\}^m \) is the target vector with $C$ classes.

\subsubsection{Chi-Square Test for Each Feature}

For each feature \( \boldsymbol{X}_i \), compute the chi-square test statistic and corresponding p-value using \eqref{eq:chi_square}:

\begin{equation}
\label{eq:chi_square}
(\chi^2_i, p_i) = \chi^2(\boldsymbol{X}_i, \boldsymbol{y}), \quad \text{for } i = 1, 2, \dots, n
\end{equation}

\subsubsection{Construct the Feature-Pair Set}

Define the set of features and their p-values as in \eqref{eq:feature_pairs}:
\begin{equation}
\label{eq:feature_pairs}
\mathcal{P} = \{(i, p_i) \mid i = 1, 2, \dots, n \}
\end{equation}

\subsubsection{Sort Features by p-Value (Ascending)}

Sort the feature indices by increasing p-value according to \eqref{eq:feature_ranking}:

\begin{equation}
\label{eq:feature_ranking}
\mathcal{P}_{\text{sorted}} = \text{sort}(\mathcal{P}) \quad \text{such that} \quad p_{i_1} \le p_{i_2} \le \dots \le p_{i_n}
\end{equation}

The resulting order places features with the strongest statistical evidence first.

\subsubsection{Select Top-$k$ Features}

Using \eqref{eq:top_k}, we select the $k$ lowest p-values and their features:
\begin{equation}
\label{eq:top_k}
\mathcal{P}_{\text{top-}k} = \left[(i_1, p_{i_1}), (i_2, p_{i_2}), \dots, (i_k, p_{i_k})\right], \quad k < n
\end{equation}

\subsubsection{Final list of features}

The selected features are given in \eqref{eq:selected_features}:
\begin{equation}
\label{eq:selected_features}
\boldsymbol{X}_{i_1}, \boldsymbol{X}_{i_2}, \dots, \boldsymbol{X}_{i_k}, \quad \text{with } p_{i_1} \le p_{i_2} \le \dots \le p_{i_k}
\end{equation}

Table~\ref{tab:feature} lists the 39 features from rank 39 to rank 1, with rank 1 corresponding to the lowest p-value.

\begin{table}[!t]
\centering
\caption{Feature Ranks Listed From 39 to 1}
\label{tab:feature}
\footnotesize
\renewcommand{\arraystretch}{1.05}
\setlength{\tabcolsep}{2.5pt}

\begin{tabular*}{\columnwidth}
{@{\extracolsep{\fill}}c l c l@{}}
\hline
\textbf{Rank} & \textbf{Feature Name} &
\textbf{Rank} & \textbf{Feature Name} \\
\hline

39 & su\_attempted & 20 & root\_shell \\
38 & src\_bytes & 19 & logged\_in \\
37 & land & 18 & flag \\
36 & wrong\_fragment & 17 & urgent \\
35 & dst\_bytes & 16 & dst\_host\_srv\_rerror\_rate \\
34 & num\_compromised & 15 & dst\_host\_rerror\_rate \\
33 & num\_root & 14 & count \\
32 & hot & 13 & serror\_rate \\
31 & num\_file\_creations & 12 & srv\_serror\_rate \\
30 & duration & 11 & rerror\_rate \\
29 & num\_access\_files & 10 & srv\_rerror\_rate \\
28 & protocol\_type & 9 & same\_srv\_rate \\
27 & srv\_count & 8 & diff\_srv\_rate \\
26 & dst\_host\_srv\_diff\_host\_rate & 7 & srv\_diff\_host\_rate \\
25 & is\_host\_login & 6 & dst\_host\_srv\_count \\
24 & dst\_host\_same\_src\_port\_rate & 5 & dst\_host\_same\_srv\_rate \\
23 & num\_failed\_logins & 4 & dst\_host\_diff\_srv\_rate \\
22 & dst\_host\_count & 3 & dst\_host\_serror\_rate \\
21 & num\_shells & 2 & dst\_host\_srv\_serror\_rate \\
 &  & 1 & is\_guest\_login \\

\hline
\end{tabular*}
\end{table}

\begin{figure*}[!t]
\centering

\subfloat[DoS attack\label{fig:bag-a}]{
\includegraphics[
    width=0.235\textwidth,
    height=4.2cm,
    keepaspectratio
]{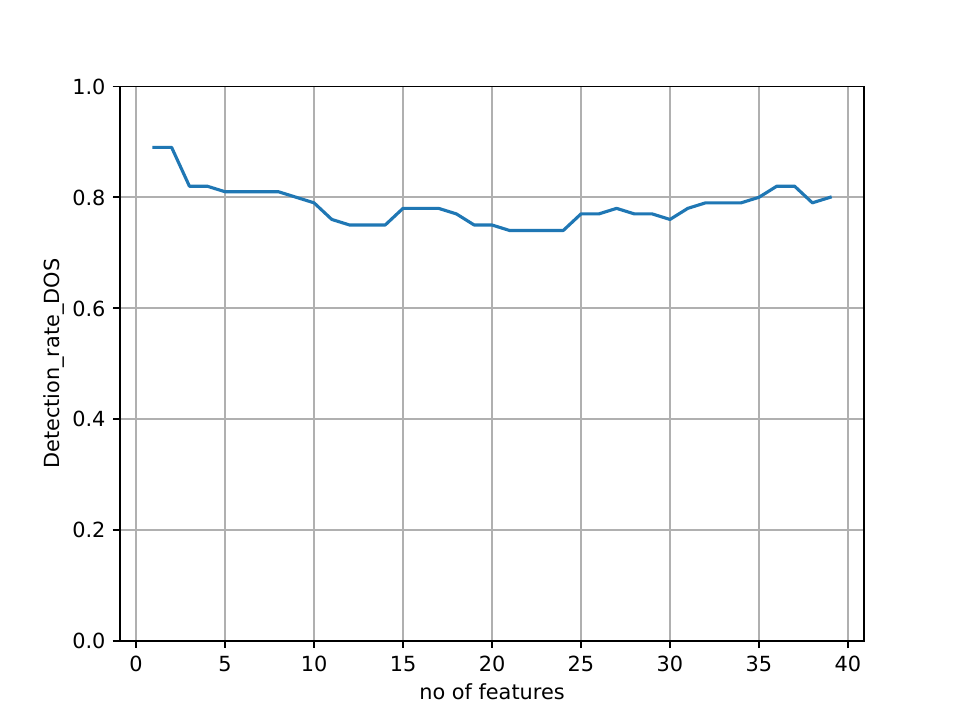}}
\hfill
\subfloat[Probe attack\label{fig:bag-b}]{
\includegraphics[
    width=0.235\textwidth,
    height=4.2cm,
    keepaspectratio
]{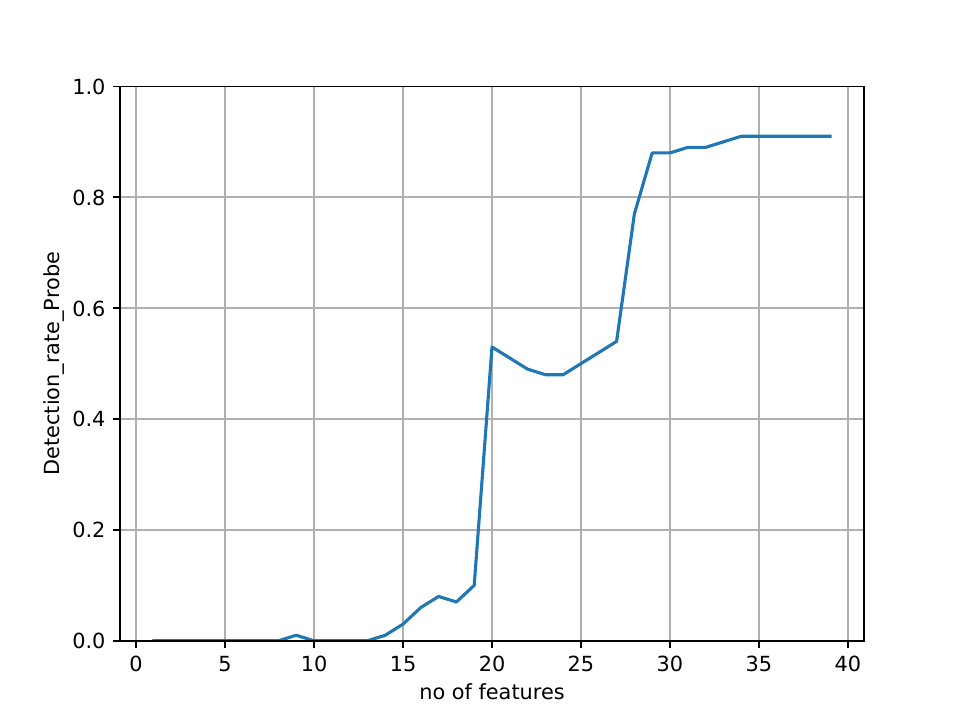}}
\hfill
\subfloat[R2L attack\label{fig:bag-c}]{
\includegraphics[
    width=0.235\textwidth,
    height=4.2cm,
    keepaspectratio
]{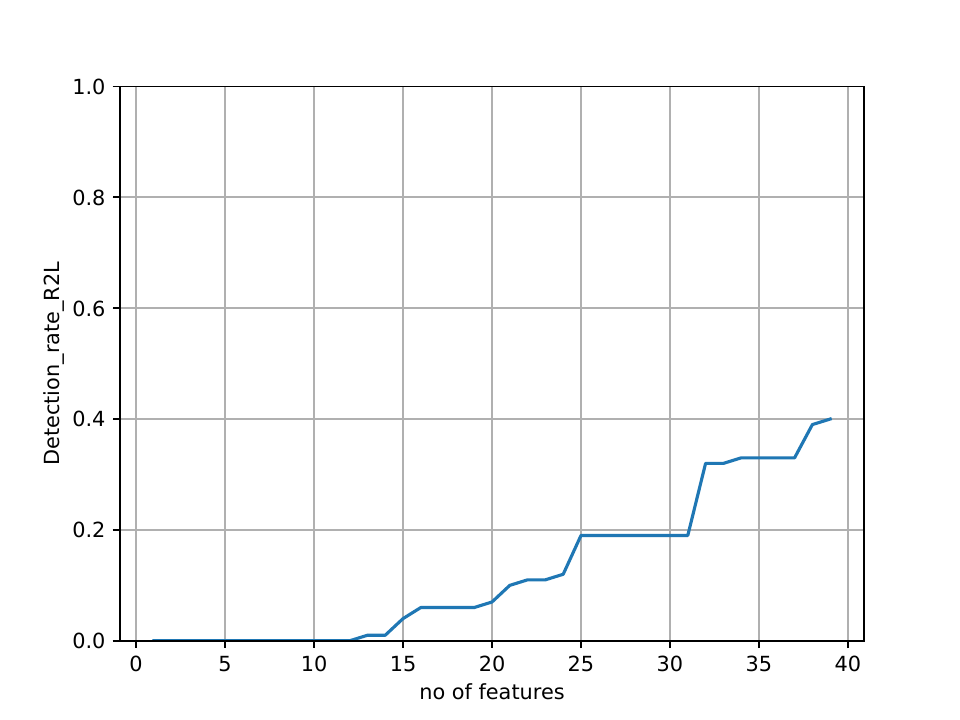}}
\hfill
\subfloat[U2R attack\label{fig:bag-d}]{
\includegraphics[
    width=0.235\textwidth,
    height=4.2cm,
    keepaspectratio
]{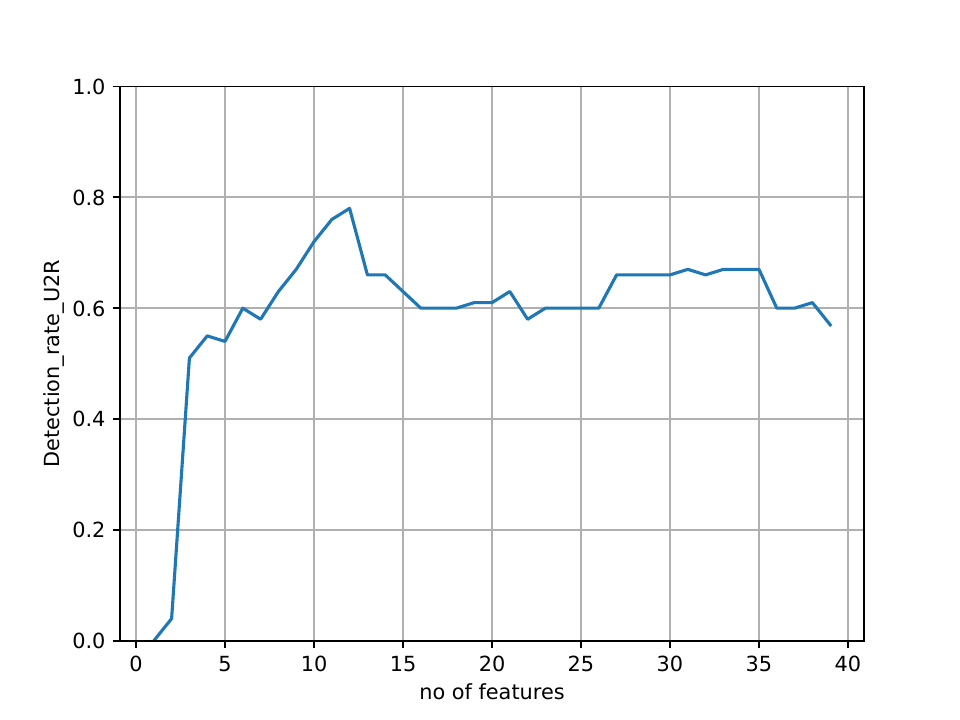}}

\caption{Detection rate (DR) by feature selection for Bagging:
(a) DoS, (b) Probe, (c) R2L, and (d) U2R attacks.}
\label{fig:bagging_results}
\end{figure*}

\begin{figure}[!htbp]
\centering{\includegraphics[width=8.2cm,height=6.8cm,keepaspectratio]{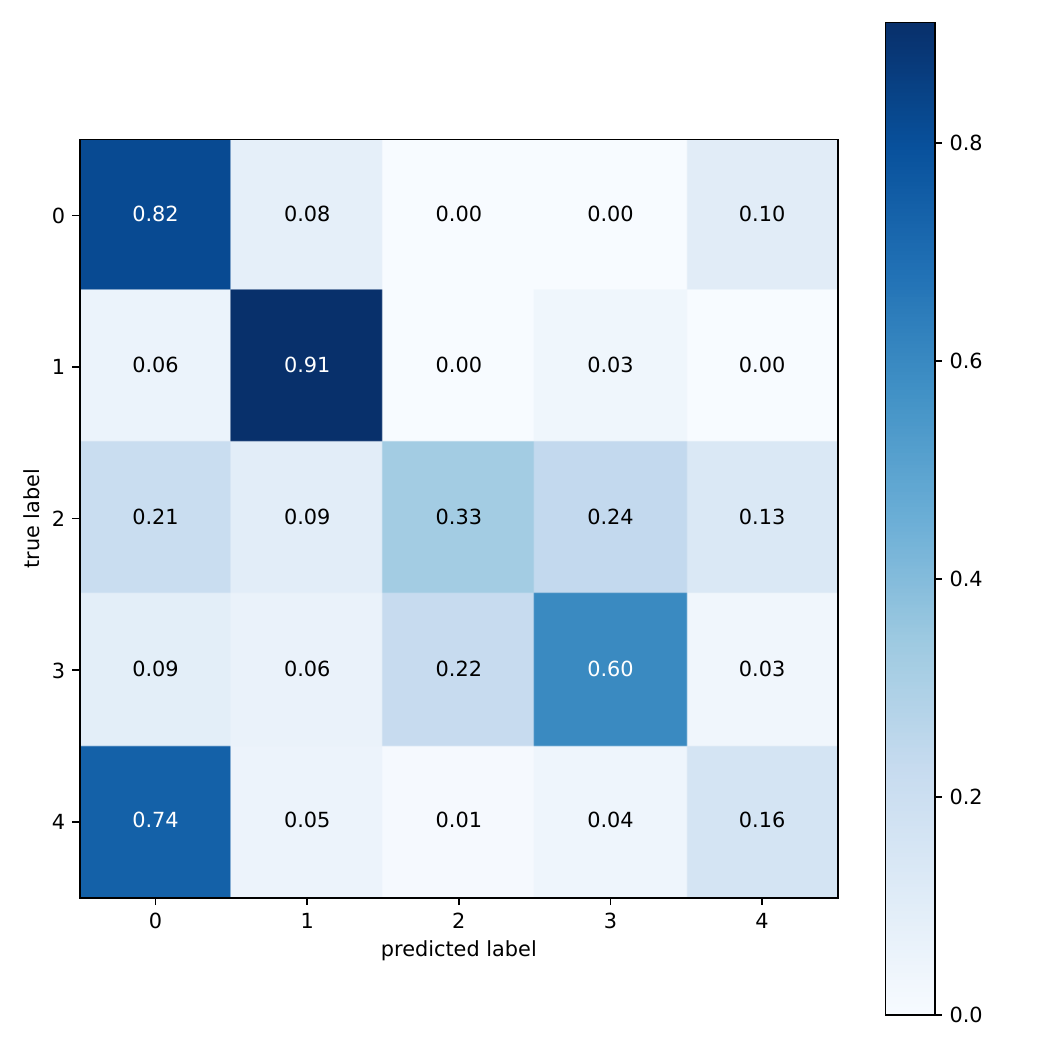}}
\caption{Confusion matrix of the Bagging\_NB classifier for normal and attack traffic. \label{cm_bagging}}
\end{figure}

\subsection{Building the Model}\label{subsec5.4.2}

We use Bagging and Boosting ensemble methods to detect network attacks in the imbalanced dataset. Naive Bayes is used as the base classifier for Bagging, while Decision Tree is used for Boosting. We develop the models in the steps below:

\begin{enumerate}
    \item Extract and prepare the dataset.
    \item Configure the base classifiers.
    \item Build the Bagging and Boosting models.
    \item Predict the target classes.
    \item Evaluate the detection rate using the confusion matrix.
\end{enumerate}

\subsection{Model Evaluation}\label{sec6}

We assess the proposed methods by examining their confusion matrices. False Negative (FN), False Positive (FP), True Negative (TN), and True Positive (TP) represent predicted results. The detection rate (DR) is obtained using these values.

\subsubsection{Detection rate}\label{subsec6.1}

Detection rate (DR), also called sensitivity or true positive rate (TPR), is the fraction of attack samples that the classifier detects correctly. It is defined in \eqref{eq:detection_rate}:

\begin{equation}
\label{eq:detection_rate}
DR = TPR = \frac{TP} {(TP+ FN)}
\end{equation}

A higher DR indicates that the model can identify a larger proportion of actual network attacks. Therefore, DR is an important performance metric for intrusion detection systems.

\begin{figure*}[!htbp]
\centering

\subfloat[DoS attack\label{fig:boost-a}]{
\includegraphics[
    width=0.235\textwidth,
    height=4.2cm,
    keepaspectratio
]{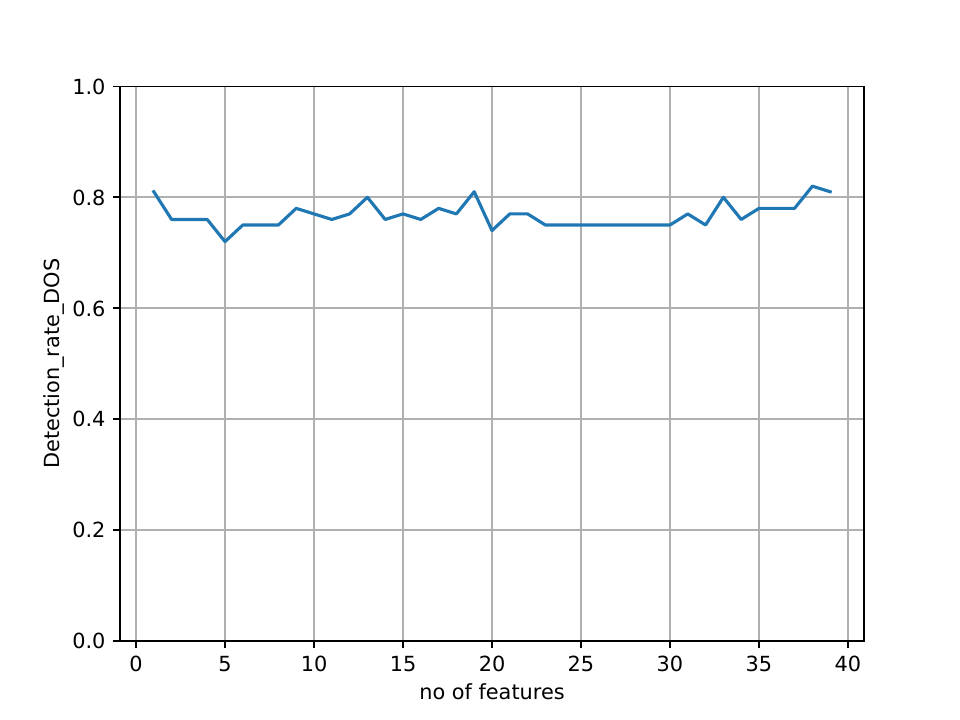}}
\hfill
\subfloat[Probe attack\label{fig:boost-b}]{
\includegraphics[
    width=0.235\textwidth,
    height=4.2cm,
    keepaspectratio
]{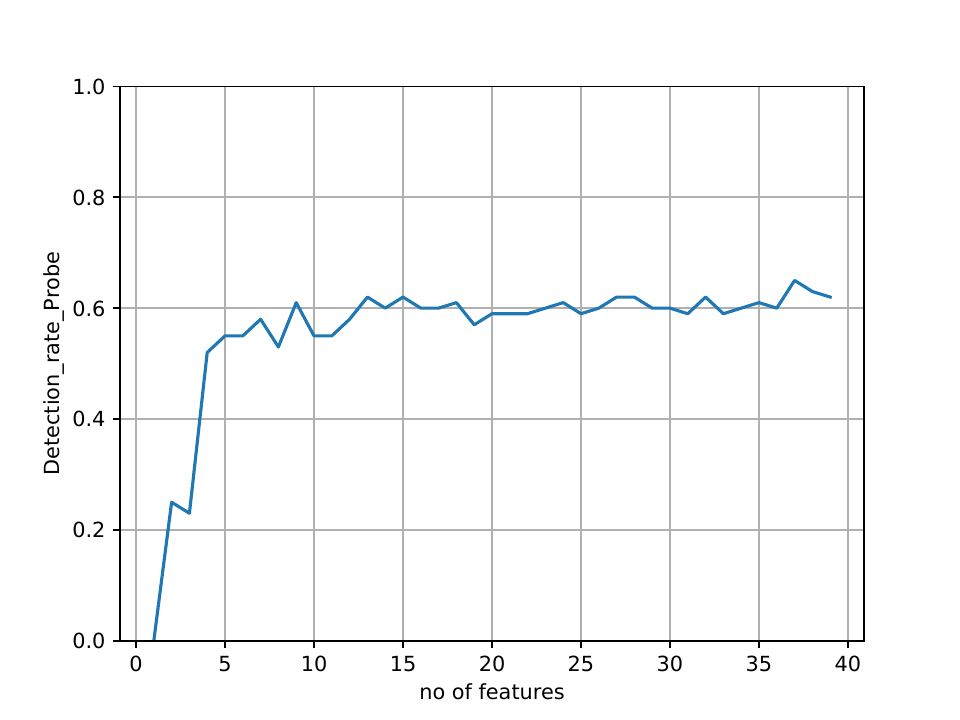}}
\hfill
\subfloat[R2L attack\label{fig:boost-c}]{
\includegraphics[
    width=0.235\textwidth,
    height=4.2cm,
    keepaspectratio
]{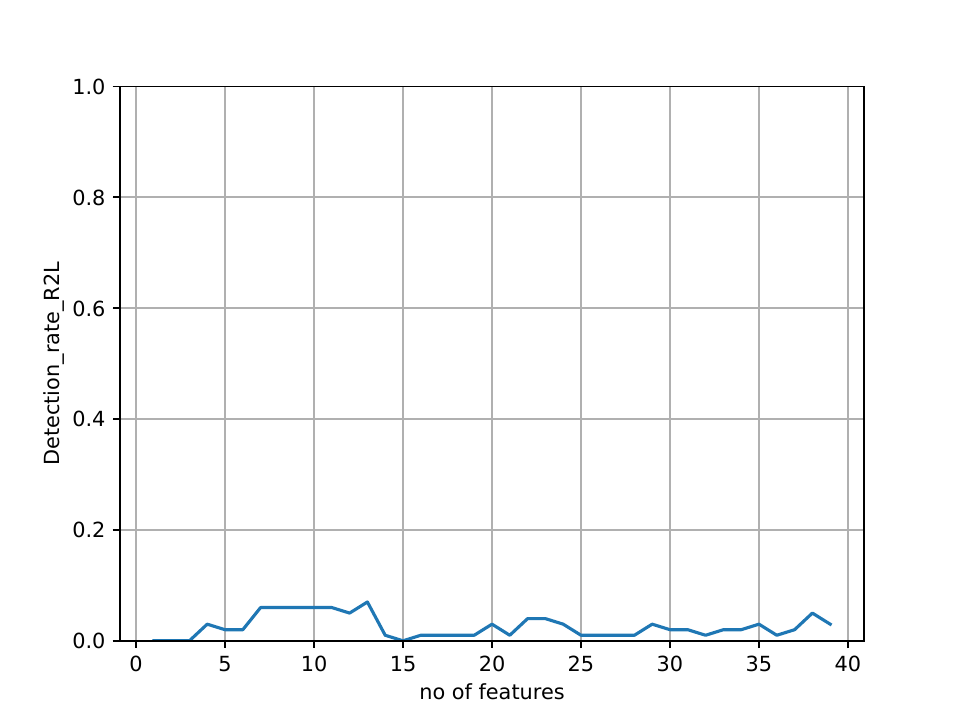}}
\hfill
\subfloat[U2R attack\label{fig:boost-d}]{
\includegraphics[
    width=0.235\textwidth,
    height=4.2cm,
    keepaspectratio
]{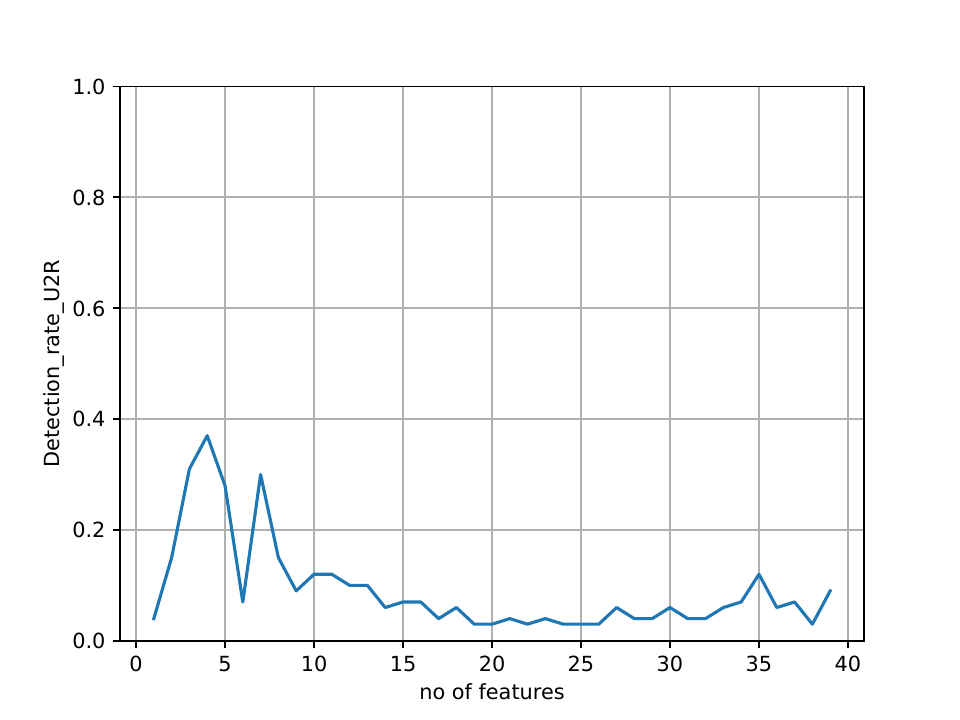}}

\caption{Detection rate (DR) by feature selection for Boosting:
(a) DoS, (b) Probe, (c) R2L, and (d) U2R attacks.}
\label{fig:boosting_results}
\end{figure*}

\subsubsection{Confusion Matrix}\label{subsec6.1.1}
A confusion matrix summarizes a model's classification outcomes. Unequal sample counts across classes can make model performance difficult to interpret accurately. This evaluation can be misleading. 

\subsubsection{Precision--Recall Curve}\label{subsec6.2}
Varying the classifier's probability threshold shows the trade-off between precision and recall. 
Recall and precision are both high when the area under the precision--recall curve is high. High precision represents few false positives, whereas high recall represents few false negatives.

\begin{algorithm}[!htbp]
\caption{Intrusion Detection With P-Value-Based Feature Selection}
\label{alg:ids_pvalue}
\small
\begin{algorithmic}[1]

\REQUIRE Network traffic dataset $\boldsymbol{\mathcal{D}}$
\ENSURE Trained model $\mathcal{M}^{*}$ and selected features
$\mathcal{X}_{k^{*}}$

\STATE Preprocess $\boldsymbol{\mathcal{D}}$ by handling missing values, encoding
categorical attributes, and normalizing numerical features.

\STATE Assign class labels as Normal, DoS, Probe, R2L, and U2R.

\STATE Split $\boldsymbol{\mathcal{D}}$ into a training set
$\boldsymbol{\mathcal{D}}_{\mathrm{train}}$ and a test set
$\boldsymbol{\mathcal{D}}_{\mathrm{test}}$.

\STATE Compute the p-value of each feature using the $\chi^2$ test
on $\boldsymbol{\mathcal{D}}_{\mathrm{train}}$.

\STATE Rank features in ascending order of p-values to obtain
$\mathcal{X}_{\mathrm{ranked}}=\{x_1,x_2,\ldots,x_n\}$.

\FOR{$k=1$ to $n$}

    \STATE Select the top-$k$ features
    $\mathcal{X}_{k}=\{x_1,x_2,\ldots,x_k\}$.

    \STATE Train classifier $\mathcal{M}_{k}$ using
    $\mathcal{X}_{k}$ on $\boldsymbol{\mathcal{D}}_{\mathrm{train}}$.

    \STATE Evaluate $\mathcal{M}_{k}$ on
    $\boldsymbol{\mathcal{D}}_{\mathrm{test}}$ and obtain $\mathrm{DR}(k)$.

\ENDFOR

\STATE Select $k^{*}=\arg\max_{k}\mathrm{DR}(k)$.

\STATE Train the candidate classifiers using $\mathcal{X}_{k^{*}}$.

\STATE Select the final model $\mathcal{M}^{*}$ based on the specified performance criterion.

\STATE Evaluate $\mathcal{M}^{*}$ using the detection rate, a confusion matrix, and a precision--recall curve.

\RETURN $\mathcal{M}^{*}$, detection rate, and $k^{*}$.
\end{algorithmic}
\end{algorithm}

\section{Experimental Results}\label{sec-result}
We implemented our proposed technique using the scikit-learn Python library. We discuss our findings below.

\subsection{Bagging\_NB}\label{subsec7.1}

From Fig.~\ref{fig:bag-a} to Fig.~\ref{fig:bag-d}, we show how the detection rates of DoS, Probe, R2L, and U2R attacks change as the number of features increases when we use the Bagging\_NB classifiers.  From Fig.~\ref{fig:bag-a}, we see that the DR is highest for DoS attacks when the number of features used is less than 4, and is second highest with 36 features. Fig.~\ref{fig:bag-b} shows the highest Probe detection rate with more than 29 features. The detection rate for R2L attacks is high when the number of features used is more than 36, as shown in Fig.~\ref{fig:bag-c}. From Fig.~\ref{fig:bag-d}, we get the highest detection rate for U2R attacks when we use 12 features and the second highest when the number of features ranges from 27 to 36.
We show the confusion matrix for the Bagging\_NB model in Fig.~\ref{cm_bagging}.

\subsection{Boosting\_DT}\label{subsec7.1.1}
\begin{figure}[!htbp]
\centering{\includegraphics[width=8cm,height=6cm]{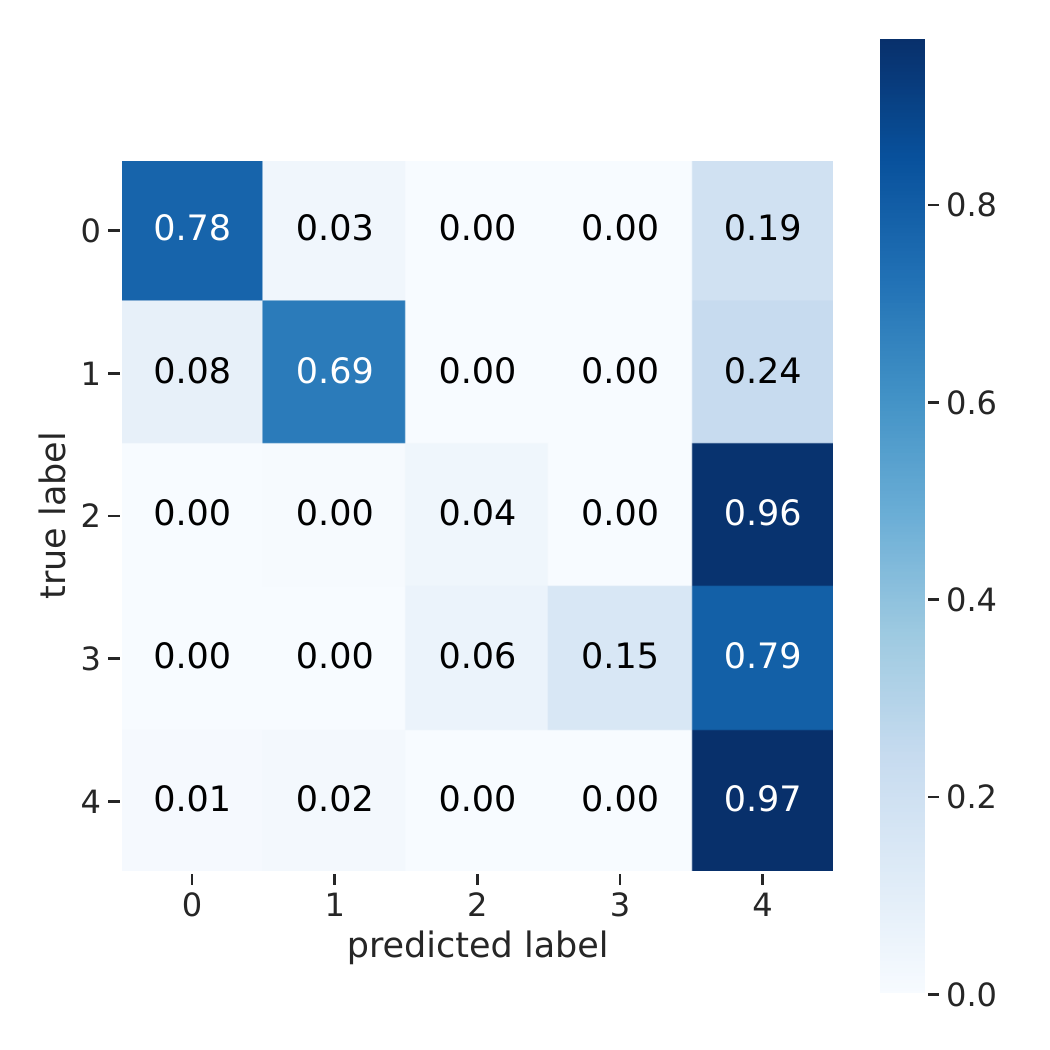}}
\caption{Confusion matrix of the Boosting\_DT classifier for normal and attack traffic.\label{cm_boost}}
\end{figure}

Figures~\ref{fig:boost-a}--\ref{fig:boost-d} show how the detection rates for DoS, Probe, R2L, and U2R attacks change as the number of features increases when using the Boosting\_DT classifier.  For DoS attacks, the highest detection rate in Fig.~\ref{fig:boost-a} occurs with 38 features. Fig.~\ref{fig:boost-b} shows the highest Probe detection rate with 37 features. Fig.~\ref{fig:boost-c} shows the highest R2L detection rate with 13 features. Fig.~\ref{fig:boost-d} shows the highest U2R detection rate with 4 features. 
Fig.~\ref{cm_boost} shows the Boosting\_DT confusion matrix.

\begin{figure}[!htbp]
\centering
\includegraphics[height=8cm,
    width=\columnwidth,
    keepaspectratio
]{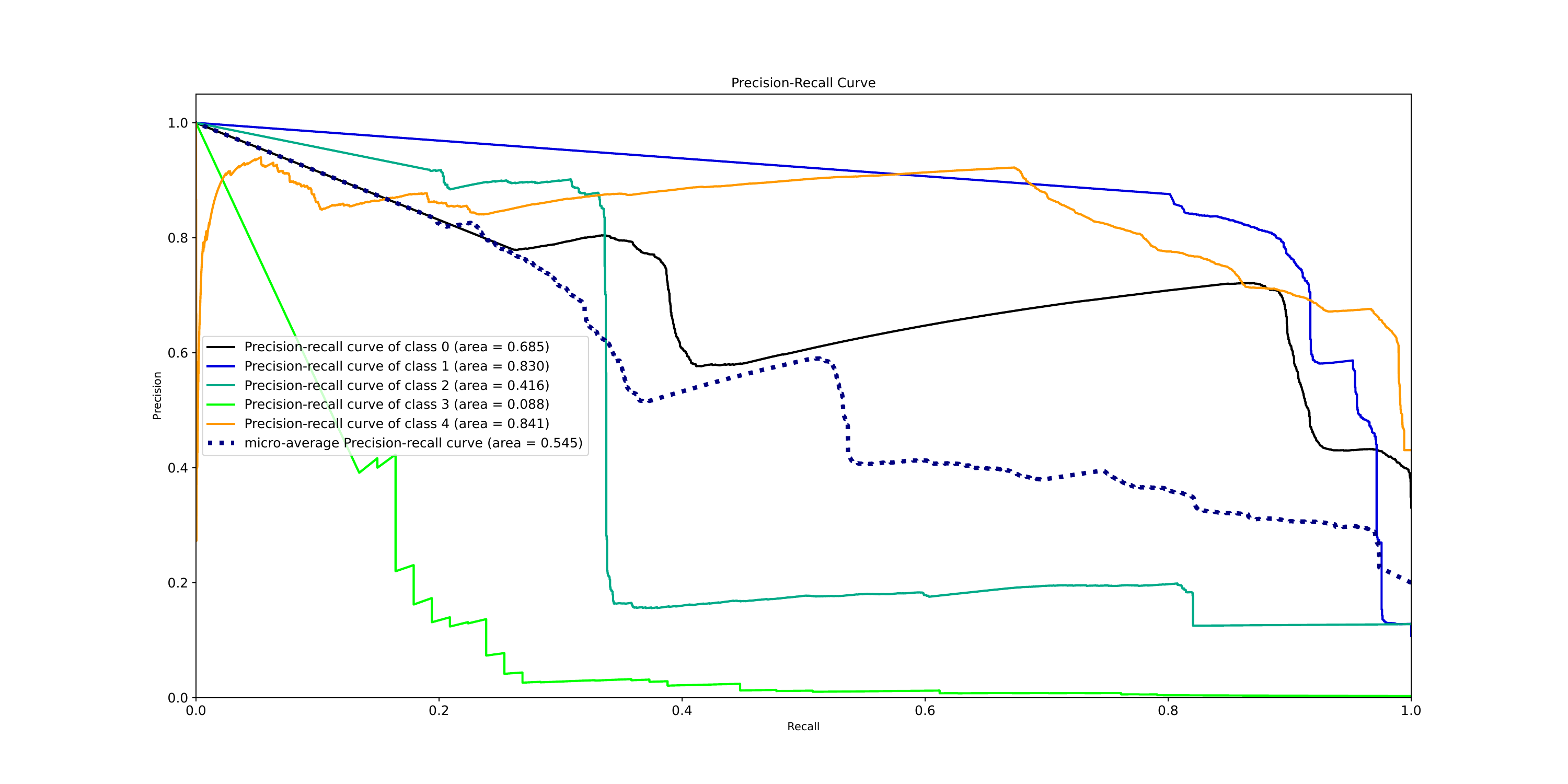}
\caption{Precision--recall curve for normal and attack traffic for the proposed models.}
\label{fig:pr}
\end{figure}







\begin{table}[!t]
\centering
\caption{Comparative Detection Rate}
\label{tab_compare}
\footnotesize
\renewcommand{\arraystretch}{1.15}
\setlength{\tabcolsep}{2.5pt}

\begin{tabular*}{\columnwidth}
{@{\extracolsep{\fill}}lcccc@{}}
\hline
\textbf{Classifier} &
\textbf{DoS (\%)} &
\textbf{Probe (\%)} &
\textbf{R2L (\%)} &
\textbf{U2R (\%)} \\
\hline

CNN \cite{ref15}       & 85 & 73 & 17 & 8  \\
DeciTree \cite{ref17}  & 84 & 66 & 21 & 7  \\
R. Forest \cite{ref17} & 81 & 68 & 2  & 1  \\
KNN \cite{ref17}       & 79 & 62 & 5  & 4  \\
LR \cite{ref17}        & 79 & 64 & 1  & 3  \\
SVM \cite{ref17}       & 77 & 48 & 9  & 2  \\
DNN \cite{ref17}       & 84 & 77 & 30 & 26 \\
Boosting\_DT   & 78 & 69 & 4  & 15  \\
\hline
\shortstack[l]{Bagging\_NB (36 features)\\{[Proposed method]}}
& 82 & 91 & 33 & 60 \\
\hline

\end{tabular*}
\end{table}


\subsection{Discussion}

From Fig.~\ref{fig:bag-a} to Fig.~\ref{fig:bag-d}, we can see that 36 features give a high detection rate for four types of attacks and normal network traffic. Fig.~\ref{cm_bagging} shows the confusion matrix using Bagging\_NB. Here, DoS, Probe, R2L, U2R and Normal traffic are represented as 0, 1, 2, 3 and 4  respectively, in Fig.~\ref{cm_bagging}.

Figure~\ref{cm_boost} shows that Boosting\_DT achieves detection rates of 78\%, 69\%, 4\%, and 15\% for DoS, Probe, R2L, and U2R attacks, respectively. Fig.~\ref{cm_boost} shows the confusion matrix using Boosting\_DT. Here, DoS, Probe, R2L, U2R and Normal traffic are represented as 0, 1, 2, 3 and 4 respectively, in Fig.~\ref{cm_boost}.

Table~\ref{tab_compare} compares Boosting\_DT and Bagging\_NB with methods reported in the literature. Other studies also evaluated DoS, Probe, R2L, U2R and Normal traffic on NSL-KDD using detection rates.
From Table~\ref{tab_compare}, we can see that the detection rates for Probe, R2L, and U2R attacks are highest for our proposed classifier. Although this classifier does not perform well in detecting normal traffic, it performs almost as well as other classifiers in the literature for DoS attacks. The proposed Bagging\_NB model achieves detection rates of 91\%, 33\%, and 60\% for Probe, R2L, and U2R attacks, respectively, exceeding the corresponding rates of the other methods listed in Table~\ref{tab_compare}.

In this paper, we also show the precision--recall curve (in Fig.~\ref{fig:pr}) of our proposed classifier which has not been shown before in the literature. Using this plot, we can see the area for successful predictions of different types of attacks individually. We can see there are a total of 5 areas (areas 0, 1, 2, 3, and 4) in Fig.~\ref{fig:pr}. Areas 0, 1, 2, 3, and 4 represent the successful prediction of U2R, DoS, R2L, Normal, and Probe traffic, respectively. The model achieves the highest area under the precision--recall curve for Probe attacks.

\section{Conclusion}\label{sec:conc}
We develop an ensemble model to identify several classes of network attacks. This model has a higher detection rate for different types of attacks. 
We evaluate Boosting\_DT and Bagging\_NB using class-specific detection rates for DoS, Probe, R2L, and U2R attacks. We conduct the experiments in scikit-learn using the NSL-KDD dataset obtained from the Canadian Institute for Cybersecurity (CIC). Our experimental workflow combines feature extraction and selection with machine-learning classification.
We obtain better detection rates in some cases than the detection rates reported in the literature when we use Bagging\_NB. For DoS attacks, the detection rate is close to the highest rate listed in Table~\ref{tab_compare}; however, for Probe, R2L, and U2R attacks, the detection rates are 91\%, 33\% and 60\% respectively, which are the highest compared with the detection rates reported in the literature.  The results indicate that Bagging\_NB is a suitable method to identify the network attacks with 36 features compared to other methods in this paper and the literature.
Future work will test alternative classifier combinations on additional network-attack datasets to investigate possible improvements in detection. We will also investigate methods to improve R2L and U2R detection, test the model on newer datasets, and measure its processing time and false-positive rate to assess its use in real-time network monitoring.


\end{document}